\documentstyle[11pt,paspconf,epsfig]{article}
\def\ltsima{$\; \buildrel < \over \sim \;$}
\def\lsim{\lower.5ex\hbox{\ltsima}}
\def\gtsima{$\; \buildrel > \over \sim \;$}
\def\gsim{\lower.5ex\hbox{\gtsima}}

\begin{document}

\title{Iron line in the afterglow: a key to the progenitor}

\author{D. Lazzati\altaffilmark{1}, G. Ghisellini and S. Campana}
\affil{Osservatorio Astronomico di Brera, via E. Bianchi 46, I-23807 
Merate (LC), Italy}

\altaffiltext{1}{Dipartimento di Fisica, Universit\`a degli studi di Milano,
via Celoria 16, I-20133 Milano, Italy}

\begin{abstract}
The discovery of a powerful and transient iron line feature in the 
X--ray afterglow spectrum of GRB~970508 and GRB~970828, if confirmed,
would be a major breakthrough for understanding the nature the progenitor
of gamma--ray bursts. We show that a large mass of iron very close  
to the burster is necessary to produce the emission line. This in 
itself strongly limits the possible progenitor of the gamma--ray event, 
suggesting the former explosion of a supernova, as predicted in the
Supranova model (Vietri \& Stella 1998). The line emission process 
and the line intensity depend strongly on the age, density and temperature
of the remnant. The simultaneous observation of the iron line and of a 
power--law optical afterglow lasted for one year strongly suggest that the 
burst emission is isotropic. Recent observations of GRB~990123 are 
also discussed.
\end{abstract}


\section{Introduction}
Piro et al. (1999) and Yoshida et al. (1999) report the detection of an 
iron emission line feature in the X--ray afterglow spectra of GRB~970508 
and GRB~970828, respectively. 
Both lines are characterized by a large flux and equivalent width (EW) 
compared with the theoretical previsions made in the framework
of the hypernova and compact merger GRB progenitor models (Ghisellini 
et al. 1999, B\"ottcher et al. 1999). The line detected in GRB~970508 is 
consistent with an iron $K_\alpha$ line redshifted to the rest--frame of 
the candidate host galaxy ($z=0.835$, Metzger et al. 1997), while GRB~970828 
has no measured redshift and the identification of the feature with the same 
line would imply a redshift $z \sim 0.33$. The line fluxes (equivalent 
widths) are $F_{Fe} = (2.8\pm1.1) \times 10^{-13}$~erg~cm$^{-2}$~s$^{-1}$ 
($\rm{EW} \sim 1$~keV) and $F_{Fe} = (1.5\pm 0.8) \times 
10^{-13}$~erg~cm$^{-2}$~s$^{-1}$ ($\rm{EW} \sim 3$~keV) for GRB~970508 and 
GRB~970828, respectively. As indicated by the flux uncertainties,
these line features are revealed at a statistical level of $\sim 3\sigma$.
A strong iron emission line unambiguously points 
towards the presence, in the vicinity of the burster, of a few per cent of
iron solar masses concentrated in a compact region. Thus the presence of 
such a line in the X--ray afterglow spectrum would represent the ``Rosetta 
Stone'' for unveiling the burst progenitor.

To date, three main classes of models have been proposed for the origin 
of gamma--ray bursts (GRB): neutron star -- neutron star (NS--NS) 
mergers (Paczy\'nski 1986; Eichler et al. 1989),
Hypernovae or failed type Ib supernovae (Woosley 1993; Paczy\'nski 1998) 
and Supranovae (Vietri \& Stella 1998).
In the NS--NS model the burst is produced during the collapse of a binary 
system composed of two neutron stars or of a neutron star and a black hole.
Interacting neutrinos or the spin--down of the remnant black hole
should power a relativistic outflow and, eventually, the production of
$\gamma$--ray photons.
Hypernovae are the final evolutionary stage of very massive
stars ($M \sim 100\, M_\odot$), whose core implodes in a black hole 
without the explosion of a classical supernova. Again, the rotational energy
of the remnant black hole is tapped by a huge magnetic field, powering the 
burst explosion. While NS--NS mergers should take place outside the original
star formation site, 
due to the large peculiar velocities and long lifetime of the coalescing 
binary systems, hypernovae take place in the dense and possibly iron rich
molecular cloud where the massive star was born.

The Supranova scenario (Vietri \& Stella 1998) assumes that,
following a supernova explosion, a fast spinning neutron star is formed,
with a mass that would be supercritical in the absence of rotation.
As radiative energy losses spin it down in a time--scale of months to years,
it inevitably collapses to a Kerr black hole, whose rotational energy can 
then power the GRB. A supernova remnant (SNR) is naturally left over around 
the burst location.

\section{Mass, size and shape of the surrounding material}

Since the detected lines are both characterized by a flux of several
$10^{-13}$~erg~s$^{-1}$, we adopt as reference a line with 
a flux\footnote{Here and in the
following we parametrise a quantity $Q$ as $Q=10^xQ_x$ and adopt cgs
units.} $F_{Fe}=10^{-13}F_{Fe,-13}$ erg cm$^{-2}$ s$^{-1}$.

This in itself constrains both the amount of line--emitting matter and
the size of the emitting region.

We assume that the emitting region is a homogeneous
spherical shell centered in the GRB progenitor, with radius $R$ and width
$\Delta R\le R$. A limit to the size of the emitting region can be set, {\it
independently from the flux variability}, imposing the flux level
times the minimum duration of the emission ($R/c$) to be less than the 
total available energy supplied by the burst fluence.
Allowing for an efficiency $q$ of the burst photon reprocessing into 
the line ($q < 0.1$, see Ghisellini et al. 1999) we have:
\begin{equation}
R<3\times 10^{17} \, q_{-1} \, \frac{{\cal F}_{-5}}{F_{Fe,-13}} \quad \hbox{cm}
\label{maxrad}
\end{equation}
where ${\cal F}$ is the total GRB fluence.
Note that this result, obtained with independent arguments, agrees with
the $\sim 1$~day variability time--scales of the detected features.

To obtain a lower limit to the total amount of mass (see Lazzati et al. 1999)
we consider a parameter $k$ given by the ratio between the total number of 
iron line photons divided by the number of iron nuclei: $k$
is the number of photons produced by a single iron atom.
This number can be constrained to be less than the total number
of ionizations an ion can undergo when illuminated by the burst
or X--ray afterglow photons. For a GRB located at $z=1$,
we obtain\footnote{The 
cosmological parameters will be set throughout this paper to
$H_0 = 65$~km~s$^{-1}$~Mpc$^{-1}$, $q_0=0.5$ and $\Lambda=0$.}: 
\begin{equation}
k\, \lsim \, {{q\, E} \over {4\pi\, \epsilon_{ion} \, R^2}} 
\, \sigma_{K} = 
6.5\times10^6 \, {{q\, E_{52}} \over {R^2_{16}}}
\label{kappa}
\end{equation}
where $E$ it the total energy of the burst photons, $\epsilon_{ion}$ the 
energy of an ionizing photon and $\sigma_K$ the ionization cross--section of 
the iron K shell. The total mass is given by the ratio of the total number of
produced photons divided by $k$ times the mass of the iron atom. We have:
\begin{equation}
M \, \gsim 0.13 \, 
{F_{Fe,-13} \; t_5\; R_{16}^2 \over q_{-1}\,A_{\odot} E_{52}} \quad 
M_\odot
\label{minmass}
\end{equation}
where $t_5$ is the time during which the line is observed in units of $10^5$
seconds and $A_\odot$ is the iron abundance in solar units.

\begin{figure}[!t]
\centerline{\epsfig{file=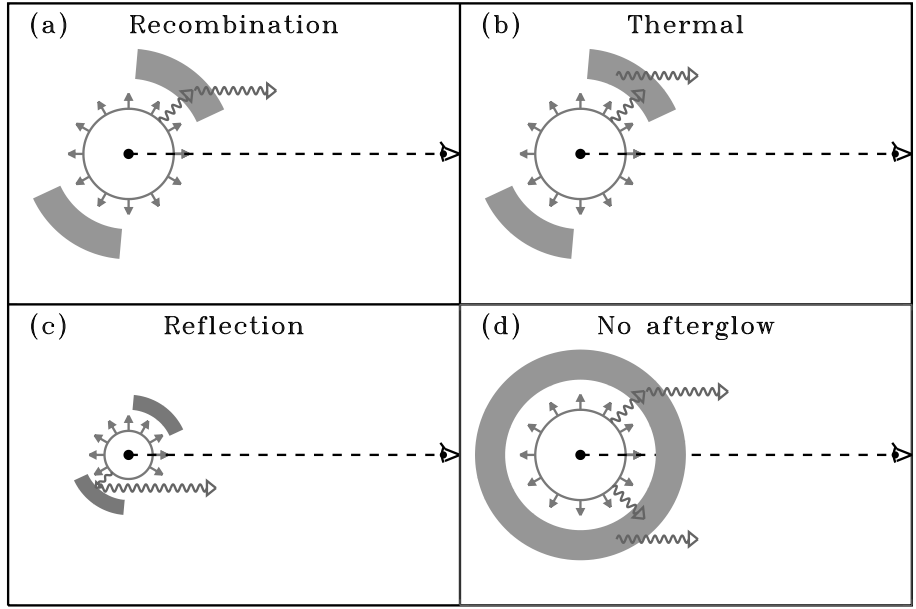,width=12cm}}
\caption{{Cartoon illustrating the different emission mechanisms discussed
in this paper. Shaded regions represent the supernova remnant, circles
(with arrows) represent the fireball. In a) photons from the bursts
photoionize iron atoms in the remnant shell, which recombine many times
during the burst. In b) the shell has been heated by the burst photons,
and emits thermally. In c) the shell, more compact than in the other cases,
produces an iron fluorescent line, and a Compton reflection continuum.
The last panel d) shows that if the remnant has a covering factor of unity,
there should be no standard afterglow emission, since the fireball is
suddenly stopped by the remnant.} \label{fig1}}
\end{figure}

If such a large amount of mass were uniformly spread around the burst
location, it would completely stop the fireball. 
In fact (see e.g. Wijers, Rees \& M\'esz\'aros 1997) the fireball is 
slowed down to sub--relativistic speeds when the swept up mass equals 
its initial rest mass. With a typical baryon load of 
$\sim 10^{-(4 \div 6)} M_\odot$, the mass predicted in Eq.~\ref{minmass} 
would stop the fireball after an observer time $t \sim \Gamma^{-2} R/c \sim
3 \times 10^4 \, \Gamma_1^{-2}$~s, i.e. almost one day.
Any surviving long wavelength emission should then decay exponentially 
in the absence of energy supply.
The only way to reconcile a monthly lasting power--law optical afterglow 
with iron line emission is through a particular geometry, 
in which the line of sight is devoid of the remnant matter.
This implies (see also Fig.~\ref{fig1}) that the burst emission is only 
moderately (if at all) beamed, since the burst photons must illuminate 
the surrounding matter and the line of sight simultaneously.

\section{Emission mechanisms}
\label{emec}

The limits on the mass discussed in the above section are very general and
independent of the mechanism producing the line photons.
Lazzati et al. 1999 give a complete analysis of three mechanisms that
are able to generate such an intense line feature.
\subsubsection{Recombination}
In this scenario, burst (or X--ray afterglow) photons keep all the iron
in the vicinity of the burst fully ionized. If the density of iron and
free electrons is large and the plasma is cool, recombination
of free electrons with iron nuclei is efficient and iron line photons
are produced. However, we require that both $n_{Fe} \gsim 10^{10}$~cm$^{-3}$
and that $T \lsim 10^4$~K. During the burst Inverse Compton interactions 
of free electrons with $\gamma$--ray photons heat the plasma to 
$T \sim 10^8$~K, and recombination is inefficient. This scenario may, 
however, be marginally consistent with GRB~970508 data if the ionization flux
is provided by an afterglow with an high energy tail but with
typical Compton temperature of $T_C \sim 10^{4 \div 5}$~K.
\subsubsection{Thermal emission}
A supernova remnant of several solar masses, some months after the explosion
of the supernova, has an emission integral of the same order of magnitude
of the intergalactic medium of a galaxy cluster. These systems emit 
$6.7$~keV iron lines due to collisional excitation of hydrogenoid
iron ions. In addition, the iron richness of a supernova remnant
can be up to ten times larger than in a cluster of galaxies, allowing for 
the emission of iron line with several keV equivalent widths.
In addition to the line, the heated SNR produces bremsstrahlung radiation
at a level $F_{ff} \sim 10^{-13}$~erg~cm$^{-2}$~s$^{-1}$,
compatible with a standard X--ray afterglow at $z \sim 1$.
\subsubsection{Reflection}
If some dense matter (e.g. a SNR with less than a month of life)
is present in close vicinity with the burster, part of the incident radiation
is reflected toward the observer. This same physical process
is efficient in the disc of Seyfert galaxies, and a strong iron line
is produced since the cross section of the iron atom is much larger than
that of free electrons. 
For a $z \sim 1$ burst, lines with up
to $F_{Fe} \sim 10^{-13}$~erg~cm$^{-2}$~s$^{-1}$ can be produced.
In steady state, $\sim 150$~eV EW lines are produced (if the material
covers half of the sky), but in gamma--ray bursts 
the reflected component is seen when the burst emission has already
faded and the EW can be much higher (even a few keV).
Given the fast variability properties of the detected lines, this model
is favored with respect to recombination and thermal emission.

\section{Progenitor}

The presence of dense matter at rest in the reference frame
of the burster (and not of the fireball) is a discriminating parameter
for the main models of burst progenitor.
Here we analyze the two main models, binary neutron star mergers 
and hypernovae, and a particular model, the Supranova, that naturally
accounts for the surrounding iron rich matter.

The production of lines in the fireball has been discussed in 
M\'esz\'aros \& Rees 1998a. They find that lines (particularly iron lines)
can be produced inside the fireball, but we should observe them at a frequency
$\nu_{obs} = \Gamma / (1+z) \nu_{l}$, where $\nu_l$ is the frequency 
of the line in the laboratory. In the case of 
GRB~970508 the observed frequency of the line is consistent with an iron line 
at the same redshift of the burster. An incredible fine--tuning is necessary
to explain the feature as an optical--UV line blueshifted in the X--ray band 
due to the fireball expansion.

The line must hence be produced by the interaction of the burst photons
with cold matter that surrounds the burster.

\subsubsection{Binary compact object mergers}
The lifetime of a binary system composed of a couple of compact
objects (NS--NS or BH--NS) ranges from 100 million to 10 billion years,
mainly depending on the initial separation and on the mass of the 
binary system components (see e.g. Lipunov et al 1995).
Due to the kick velocity that a compact object inherits
from the supernova explosion that generated it, $\sim 200$~km~s$^{-1}$ 
on average (Kalogera et al. 1998), these binary systems merge well outside 
their formation site and, possibly, even outside the host galaxy.
Assuming an hydrogen density $n = 1$~cm$^{-3}$ and a solar
iron abundance, a sphere of radius:
\begin{equation}
R \gsim 3.7 \times 10^{23} \, n^{-1} \, {F_{Fe,-13} \, t_5 
\over q_{-1} \, A_\odot \, E_{52}} \qquad {\rm cm}
\label{radio}
\end{equation}
is necessary to meet the requirements of Eq.~\ref{minmass}.
However, the radius of the above equation is five orders of magnitude
larger than the maximum allowed value (cf. Eq.~\ref{maxrad}).

\subsubsection{Hypernova}
If we insert in Eq.~\ref{radio} the density appropriate for a molecular
cloud, in which a very massive star is thought to be located,
we obtain $R \sim 3 \times 10^{18} A_\odot^{-1}$~cm for $n = 10^5$~cm$^{-3}$
that, for an iron rich cloud, is marginally consistent with the limit
of Eq.~\ref{maxrad}. To firmly rule out the possibility of an
intense iron line produced in the hypernova scenario, we analyze
each emission mechanism discussed in Sect.~\ref{emec}.
Reflection can be immediately excluded since a radius $R>10^{20}$~cm
cloud is necessary to have a Thomson thick mirror. The efficiency of
thermal emission is proportional to the square of the density.
Lazzati et al. 1999 estimate that both a density $n \sim 10^{10}$~cm$^{-3}$ 
and a mass of several solar masses (see also Sect.~\ref{emec})
are necessary to produce such an intense iron line. In the hypernova
scenario, the density is 5 orders of magnitude smaller, giving
a flux 10 orders of magnitude fainter.
The recombination time in a medium with the considered
density is too long (see B\"ottcher et al. 1999), however a fluorescence line
may be emitted due to ionization of the inner shell of almost neutral 
iron atoms. The more optimistic estimate of the line flux produced by 
fluorescence in a molecular cloud is given in Ghisellini et al. 1999,
with an upper limit $F_{Fe} < 5 \times 10^{-15}$~erg~cm$^{-2}$~s$^{-1}$.
Even if we consider the presence of a pre--hypernova wind with $\dot m \sim
10^{-4} M_\odot$/yr, the total mass contained in a radius
of $10^{18}$~cm would be a few percent of a solar mass, leaving the 
above discussion unaffected.
We hence conclude that in a standard hypernova scenario it is not possible
to produce an intense iron line.

M\'esz\'aros \& Rees 1998b discussed the possible production of an
intense iron line if a compact companion of the hypernova has left
a dense torus of material ($M \sim M_\odot$; $n \sim 
10^{10}$~cm$^{-3}$) at a distance $R \sim 10^{15}$ cm from the burster.
Another progenitor model can account for this matter in a more natural way.

\section{Supranova}
In the model presented by Vietri \& Stella 1998 the burst explosion
is preceded by a supernova. 
The neutron star left over by the supernova explosion is supermassive, i.e.
its mass exceeds the limiting mass for a neutron star, but the implosion
in a black hole is prevented by the fast rotation.
As the system spins down, due to radiative
losses, the neutron star shrinks until a limiting state is reached and
the star collapses in a black hole. The duration of this transient phase 
depends from the strength of the magnetic field embedded in the star and from 
the initial spin velocity, but should typically last between several months and
a dozen of years. 
If we consider a supernova exploded a year before the burst, the density 
of its shell remnant at the burst onset is:
\begin{equation}
n \sim 5\times10^9 \, M_\odot \, v_9^3 \quad {\rm cm}^{-3}
\end{equation}
where $M_\odot$ is the mass of the ejecta in solar units, $v_9$ is
their speed in units of $10^9$~cm~s$^{-1}$ and the width of the
shell has been assumed $\Delta R = R/100$.
The shell radius is $R = 3\times 10^{16}$~cm, in good agreement with 
the constraints derived above. If we consider that a SNR can have
an iron richness ten times the solar value (Woosley 1988), this scenario
is the most natural one to explain the intense iron line observed in the
afterglow of GRB~970508 and GRB~970828.

\section{GRB~990123}

The recent explosion of GRB~990123 has gathered the attention of the 
gamma--ray astronomers due to the huge fluence and large distance
of this burst. The early X--ray afterglow flux was measured by the 
{\it Beppo}-SAX satellite at a level of $F \sim 1.1 \times 
10^{-11}$~erg~s$^{-1}$ (Heise et al. 1999) about 6 hours after
the burst trigger. The first results on the spectral analysis
gives a ``featureless spectrum'' (Heise et al. 1999).
If an iron line feature with the same EW of those of GRB~970508 and 
GRB~970828 were present in the afterglow of GRB~990123,
a firm detection would have been established: since the underlying
continuum flux of the latter burst is $\sim 100$ times larger than in
the other two cases, the signal to noise ratio of the line would have been
10 times larger. Moreover, due to the larger redshift of GRB~990123
(z=1.6004, Kulkarni et al. 1999) the redshifted line energy lies in a 
region where the detector efficiency is the largest.
This could have led to a $30\,\sigma$ line feature.

There are, however, two reasons why an iron line is not
expected in the afterglow of GRB~990123.
First, (Lazzati et al. 1999) the line intensity is 
limited by the recombination time and not by the ionization time.
This means that, even in the case of GRB~970508 - a weak burst -
the ionizing flux was more than enough to fully exploit the
matter surrounding the progenitor.
If an iron line with flux $F_{Fe} \sim 3 \times 
10^{-13}$~erg~cm$^{-2}$~s$^{-1}$ was present in the X--ray afterglow
of the 23 January burst, its EW would be decreased to the very low
value of $\sim 10$~eV, hence well below the detection threshold
of the {\it Beppo}-SAX.
Finally, the break in the power--law decrease of the optical light curve
of GRB~990123 afterglow has been interpreted as a sign of beaming in the
burst itself (Kulkarni et al. 1999; Fruchter et al 1999). 
If this interpretation is correct, even if a cloud
were present in close vicinity with the burster, its illumination would 
have been minimal, preventing the formation of an intense line.

\section{Discussion}

We have analyzed how the detection of an iron line redshifted to the rest 
frame of the burster can be used to constrain the progenitor models.
Indeed, due to the relativistic outflows involved in the
fireball model, very poor information of the progenitor is contained
in the gamma--ray burst photons, and the only mean to investigate
the progenitor nature is through the interaction of the photons with 
the ambient medium.

In particular, we have analyzed three different models predicting
very different conditions of the external medium: the merging
of two compact objects in a binary system, the hypernova and the Supranova.
We derived a (model--independent) lower limit to the mass required to produce 
the line, ruling out the NS--NS merger model and, with a careful analysis, 
the hypernova.
The Supranova, which naturally accounts for a shell remnant
of iron rich material surrounding the burster seems the most 
promising model capable to produce an intense iron line.

Due to the paucity of the statistical significance of the line
detections, more observations are needed to draw firm conclusions.
We show that, despite its luminosity, GRB~990123 is not the ideal
test for the line detection and more observations with more efficient 
satellites are needed.

\acknowledgments

DL thanks the Cariplo foundation for financial support.


\begin{references}
\reference B\"ottcher, M., Dermer, C. D., Crider, A. W., \& Liang, E. P.
	1999, \aap, 343, 111
\reference Eichler, D., Livio, M., Piran, T., \& Shramm, D. N. 1989, Nature, 
	340, 126
\reference Fruchter A. S., et al. 1999, ApJ submitted (astro-ph/9902236)
\reference Heise, J., et al.
1999, IAUC n. 7099
\reference Kalogera, V., Kolb, U., \& King, A. R. 1998, \apj, 504, 967
\reference Kulkarni, S., et al. 1999, Nature, 398, 389
\reference Lazzati, D., Campana, S., \& Ghisellini, G. 1999, \mnras, 304, L31
\reference Lipunov, V. M., Postnov, K. A., Prokhorov, M. E., \&
	Panchenlo, I. E. 1995, \apj, 454, 593
\reference M\'esz\'aros, P., \& Rees, M. J. 1998a, \apj, 502, L105
\reference M\'esz\'aros, P., \& Rees, M. J. 1998b, \mnras, 299, L10
\reference Metzger, M. R., et al. 1997, Nature, 387, 878
\reference Paczy\'nski, B. 1986, \apj, 308, L43
\reference Paczy\'nski, B. 1998, \apj, 494, L45
\reference Piro, L., et al. 1999, \apj, 514, L73
\reference Vietri, M., \& Stella. L. 1998, \apj, 507, L45
\reference Wijers, R. A. M. J., Rees, M. J., \& M\'esz\'aros, P. 1997, \mnras, 
	288, L51
\reference Woosley, S. E. 1988, \apj, 330, 218
\reference Woosley, S. E. 1993, \apj, 405, 273
\reference Yoshida, A., et al.
1999, \apj~in press
\end{references}
\end{document}